\magnification=1200
\input amstex
\documentstyle{amsppt}
\voffset=-3pc
\loadbold
\loadmsbm
\loadeufm

\def\bC{\Bbb C}

\def\bR{\Bbb R}

\def\mt{\mapsto}
\def\ra{\rightarrow}

\def\pr{\text{pr}}

\def\Vect{\text{Vect}}

\def\ad{\text{\rm ad}}
\def\sh{\text{\rm sh}}
\def\ch{\text{\rm ch}}
\topmatter
\title{Curvature of fields of quantum Hilbert spaces}
\endtitle
\author L\'aszl\'o Lempert and R\'obert Sz\H{o}ke$^{*)}$
\footnote""{$^*$Research partially supported by NSF grant DMS 0700281
and  OTKA grants
 N81203.} 
\endauthor
\rightheadtext{L\'aszl\'o Lempert and R\'obert Sz\H{o}ke }
\leftheadtext{Curvature of fields of quantum Hilbert spaces}
\address
\hskip-.20truein Dept.~of Mathematics, Purdue University, West Lafayette, 
IN\ 47907, USA\newline
Dept.~of Analysis, Institute of Mathematics, E\"otv\"os University, 
P\'azm\'any P.~s\'et\'any 1/c, Budapest 1117, Hungary\endaddress
\abstract
We show that using the family of adapted K\"ahler polarizations of the phase
space of  a compact, simply connected,  Riemannian 
symmetric space of rank-1, the obtained field $H^{corr}$
of quantum Hilbert spaces produced by geometric quantization including the 
half-form correction 
  is flat  if $M$ is the 3-dimensional sphere and not even
projectively flat otherwise.

\endabstract
\endtopmatter
\document
\subhead 1. Introduction
\endsubhead
Suppose an $m$--dimensional compact Riemannian manifold $M$ is the classical 
configuration space of a mechanical system, the metric corresponding to 
twice the kinetic energy.
To quantize it according to the prescriptions of Kostant and Souriau 
[Ko,So,Wo], one first passes to phase space $N$, which for the moment is 
taken $TM\approx T^*M$, a symplectic manifold with an exact symplectic form 
$\omega$, equal to $\sum dq_j\wedge dp_j$ in the usual local coordinates.
The prequantum line bundle is a Hermitian line bundle $E\to N$ with a 
connection whose curvature is $-i\omega$.
If $M$ is simply connected, the bundle is unique up to a connection 
preserving Hermitian isomorphism.
In any case, one such line bundle is obtained from a real 1--form $a$ on $N$ 
such that $da=-\omega$, by letting $E=N\times\bC\to N$ to be the trivial 
line bundle with $h^E(x,\gamma)=|\gamma|^2$ the trivial metric on it. 
If sections are
identified with functions $\psi\colon N\to\bC$,
the connection $\nabla^E$ is defined by
$$
\nabla_\zeta^E\psi=\zeta\psi+ia(\zeta)\psi,\qquad\zeta\in\Vect \,N.
$$
A choice of a  K\"ahler  structure on $N$  with K\"ahler form $\omega$ induces
on $E$ the structure of a 
holomorphic line bundle. This gives rise to the quantum Hilbert space $H$,
consisting of holomorphic sections of $E$ that are $L^2$ with
respect to the volume form $\omega^m/m!\,$.

Often one is forced to include in this 
construction the so called half-form correction. Suppose
 $\kappa$ is a square root of the canonical
 bundle $K_N$. Then the corrected quantum Hilbert space $H^{corr}$ consists of 
the $L^2$ holomorphic sections of $E\otimes\kappa$.

 When $M$ is a real-analytic Riemannian manifold, there is  a natural 
K\"ahler polarization on (some subset of) $N$.
In [Sz1,GS] the second author and Guillemin--Stenzel construct a 
canonical complex structure (``adapted complex structure'' or 
``Grauert tube'') on a neighborhood $X\subset TM$ of the zero section, 
in which $\omega$ becomes a K\"ahler form (see also [H-K]). 
 In good cases $X=N$. One gets examples of this sort  when
 $M$ is a compact normal
Riemannian homogeneous space,   
but there are nonhomogeneous examples as well, see [A,Sz1, Sz2].

In fact the adapted complex structure is just one member
 in a natural family of K\"ahler structures on $N$ [L-Sz2].
To see this it is advantageous to adhere to Souriau's philosophy ([So]) and 
define
the phase space $N$ of a compact Riemannian manifold not as $TM$ or $T^*M$ but
 as the manifold of parametrized geodesics $x:\bR\ra M$. Any $t_0\in\bR$ 
induces a diffeomorphism $N\ni x\mt\dot x(t_0)\in TM$, and  the pull back of 
the  canonical 
symplectic form of $TM\approx T^*M$ is independent of $t_0$; we denote it by 
$\omega$. We  identify $M$ with the submanifold of zero speed geodesics in $N$.
Affine reparametrizations $t\mt a+bt$, $a, b\in\bR$, act on $N$ and define
a right action of the Lie semigroup $\Sigma$ of affine reparametrizations.  

  Given a complex manifold structure on $\Sigma$, 
a complex structure on $N$ 
is called adapted if for every $x\in N$ the orbit map 
$\Sigma\ni\sigma\mt x\sigma
\in N$ is holomorphic ([L-Sz2]). 
An adapted complex structure on $N$ can exist only if
the initial compex structure on $\Sigma$ is left invariant. Left invariant 
complex structures on $\Sigma$ are parametrized by the points of 
$\bC\setminus\bR$. For each $s\in \bC\setminus\bR$ and corresponding left
 invariant complex structure $I(s)$ on $\Sigma$, if an $I(s)$ adapted complex 
structure $J(s)$ exists on $N$, 
then this structure is unique and if $J(i)$ exists, then $J(s)$ also exists
 for all $s$ in $s\in \bC\setminus\bR$. The points of the
upper half plane (denoted from now on by $S$) correspond to $J(s)$ 
in which $\omega$ is a Kahler form.
The original definition
 of adapted complex structures in [L-Sz1, Sz1] corresponds to the parameter 
$s=i$.

Now suppose  for the compact Riemannian manifold $M$ the adapted complex
structure $J(i)$ exists on $N$. 
 With the help of the corresponding   family of K\"ahler structures $J(s)$
on $N$, 
geometric quantization
produces a family $H_s$ of quantum Hilbert spaces. Our main concern is how (and
 when) can one define a natural (projective) isomorphism among these Hilbert
spaces.

To deal with this problem, a key idea, following [ADW] and [Hi], is that 
   the collection $\{H_s : s\in S\}$ resembles a holomorphic Hermitian vector 
bundle, in which one can try to construct a Chern-like canonical 
connection, and use its parallel transport canonically to identify the 
different fibers $H_s$. To what extent this can be done was explored in [L-Sz3].
The starting point is that  the
family of adapted complex structures $J(s)$, $s\in S$ on $N$ can all 
be put together to form a holomorphic fibration $\pi:Y\ra S$; where 
the fibers
 $Y_s=\pi^{-1}s$ are biholomorphic to  $(N,J(s)).$ 
In fact, as a differentiable manifold, $Y=S\times N$, and the projection
pr$:Y\ra N$ realizes the biholomorphisms $Y_s\ra(N,J(s)),$
([L-Sz2, Theorem 5]). 

 Armed with this fibration one can perform geometric quantization
simultaniously. 
As we shall see shortly, 
the object we get is what we call a field of Hilbert spaces.  
A field of Hilbert spaces is simply a map $p:H\rightarrow S$ of sets with each
fiber $H_s=p^{-1}(s)$ endowed with the structure of a Hilbert space.
When $S$ is a smooth or real analytic manifold, one can introduce the notion
of a smooth or analytic structure on $p:H\ra S$, by specifying a set
$\Gamma^\infty$ resp. $\Gamma^\omega$ of sections of $p$, together with operators
$\nabla_\xi:\Gamma^\infty\ra\Gamma^\infty$ for all vector fields $\xi$ on $S$. The
set $\Gamma^\infty$ and the operators $\nabla_\xi$ are supposed to satisfy 
certain axioms (see [L-Sz3, Sect.2]).

Smooth fields of Hilbert spaces are looser structures than Hilbert bundles,
but the notion  is strong enough to define
  curvature, (projective) flatness and (local) triviality of the 
field.  
In particular a  smooth Hilbert field $H\rightarrow S$  is
called projectively flat if the curvature operator 
$$
R(\xi,\eta)=\nabla_\xi\nabla_\eta-\nabla_\eta\nabla_\xi-\nabla_{[\xi,\eta]}:
\Gamma^\infty\rightarrow\Gamma^\infty
$$ 
is multiplication by a 
function $r(\xi,\eta):S\rightarrow\bC$. Just like  with vector bundles, 
$r$ is in fact a smooth closed 2-form on $S$, and  a simple
twisting will reduce projectively flat smooth Hilbert fields to flat ones.
Flatness and projective flatness are important, because in a flat and analytic
field a parallel transport can be introduced that identifies the fibers
canonically. Similarly, in   projectively flat analytic fields the corresponding
parallel transport identifies the projectivized fibers ([L-Sz3, Theorem 2.3.2,
Theorem 2.4.2]).

Now back to geometric quantization of a compact Riemannian manifold $M$,
assuming the adapted complex structure $J(s)$ exists on the entire manifold
$N$ of geodesics.
To quantize $(N,J(s))$ simultaneously, construct a Hermitian holomorphic
line bundle $E\ra Y$ with curvature $-i\tilde\omega:=-i\pr^*\omega$. The 
restriction of $E$ to $Y_s$ yields the prequantum line bundle corresponding
to  $(N,J(s),\omega)$. The restriction  of the form $\nu=\tilde\omega^m/m!$
to a fiber $Y_s$ is a volume form.
 The spaces of holomorphic
$L^2$-sections of $E|Y_s$ form the  Hilbert field 
$H\ra S$. 
Assuming now that $M$ is simply connected, there is a unique Hermitian 
holomorphic line bundle $\kappa$ on $Y$, so that 
$\kappa\otimes\kappa\approx K_\pi$ (the relative canonical bundle of $Y$ with 
 $K_\pi|Y_s$ being the canonical bundle of $Y_s$). The spaces of holomorphic
$L^2$-sections of $E\otimes\kappa|Y_s$ form the corrected Hilbert field 
$H^{corr}\ra S$.

More generally Hilbert fields naturally arise as direct images of 
holomorphic vector bundles.
Suppose $\pi:Y\rightarrow S$ is a surjective holomorphic submersion of
 finite dimensional complex manifolds, not necessarily proper. 
Let $\nu$ be a smooth form on $Y$ that restricts to a volume form
on each fiber $Y_s=\pi^{-1}s$ and let $(E,h^H)\rightarrow Y$ be a Hermitian  
holomorphic vector bundle of finite rank. 
Let $H_s$ be the Hilbert space of $L^2$ holomorphic sections of $E|Y_s$. 
 The spaces $H_s$ form a Hilbert field $H\rightarrow S$. 

Under certain conditions on $Y$ and $E$, the field comes naturally 
 endowed with a smooth
 structure ([L-Sz3, Sect. 6, 7]). In the problem of geometric quantization by
adapted complex structures, these conditions are known to be satisfied 
in the special case when 
 $M$ is a 
compact, simply connected, normal 
Riemannian homogeneous space. In fact, in this case
$H^{corr}\ra S$
turns out to be analytic ([L-Sz3, Theorem 11.1.1]).

 Our main result is the following:
\proclaim{Theorem 1.1} Let $M$ be a compact, simply connected,  Riemannian 
symmetric space of rank-1. Then the corresponding field $H^{corr}$
of quantum Hilbert spaces
  is flat  if $M$ is the 3-dimensional sphere and not even
projectively flat otherwise.
\endproclaim

We prove this result in Sect.5. It shows  quantization is unique 
for the $3$-sphere  and in the rest of the cases quantization 
 does depend on the choice
of the K\"ahler polarization.

Flatness also implies  $H^{corr}\ra S$ is a genuine 
 Hilbert bundle (trivial in this case),
 something that is not known to be true for the other rank-1 symmetric 
spaces.

The situation for the higher rank symmetric spaces is  more complicated
 and will be 
treated in a separate publication \cite{L-Sz4}.

\subhead 2. Curvature calculations
\endsubhead

Consider a simply connected, compact, Riemannian symmetric space
 $(M^m,g)$ and $H^{corr}\ra S$ the corresponding field of quantum Hilbert spaces.
Let $U$ denote the identity component
 of the  isometry group of $M$ 
and $K\subset U$ the isotropy group of a fixed $o\in M$.
Let $\frak u$ and $\frak k$ be the Lie algebras of $U$ and $K$
and let $\frak p_*\subset\frak u$ be the orthogonal complement of $\frak k$.

 $U$ acts on $(N,J(i))$ by biholomorphisms and this action 
 induces a representation $\hat \pi$ on $\Cal O(N, J(i))$, by the formula
$av=(a^{-1})^*v$ (pull back by $a^{-1}$), where $a\in U$, 
$v\in \Cal O(N, J(i))$. 
The same formula defines a unitary representation $\pi$ on $L^2(M)$. 
The restrictions $V_\chi|_M$ of the isotypical subspaces of   $\hat \pi$  are
precisely the isotypical subspaces of $\pi$ and the latter are well known to be
finite dimensional.
Since $M$ is a  maximal dimensional, totally real submanifold 
in $N$, we get that $V_\chi$
  are also finite dimensional.
 The isotypical subspaces of $\pi$ are parametrized by the 
irreducible spherical (w.r.t. $K$) representations of $U$ ([He2, Theorem 4.3]).
 In fact the restrictions of $\hat\pi$ to the isotypical subspaces
 $V_\chi$ (or equivalently the restrictions
 of $\pi$ to $V_\chi|_M$) are precisely these spherical
representations.

Flatness of the field $H^{corr}\ra S$ can be understood in terms of certain
operators $P_\chi(s)$ on $V_\chi$. Namely $H^{corr}\ra S$ is flat (resp. 
projectively flat) if and only if $P_\chi(s)$ are of the form 
$P_\chi(s)=p_\chi(s)Id_\chi$ 
and $\bar\partial\partial\log p_\chi(s)=0$ for all $\chi$ (resp.  
$\bar\partial\partial\log p_\chi(s)$ is independent of $\chi$), see 
[L-Sz3, Theorem 9.2.1].
 According to [L-Sz3, Lemma 11.2.1] and [L-Sz3, Sect. 12.1],

$$
p_\chi(s)=\frac{c_\chi }{(\text{Im }s)^{m/2}}
\int\limits_{\frak p_*}\int\limits_K 
e^{-\frac{|\zeta|^2}{\text{Im }s}}\chi (k\exp (-2i\zeta))
\sqrt{\eta(\zeta)}
dk \,d\zeta,\tag2-1
$$
where $c_\chi$ is independent of $s$,
$dk$ is normalized Haar measure on $K$, $d\zeta$ translation
invariant Lebesgue measure
on $\frak p_*$, and

$$
\eta(\zeta):=\det \left(\left.\frac{\sin 2\ad\zeta}{\ad\zeta}
\right|_{\bC\otimes\frak p_*}\right).
$$

The function $f_\chi(g)=\int_K\chi(k g^{-1})dk$ , occuring in (2-1), is known as 
spherical function, corresponding to the character $\chi$, see 
\cite{He2, IV., Theorem 4.2}. This function has a  holomorphic extension to
 the complexified group $U_\bC$ that we also denote by $f_\chi$. 

\proclaim{Proposition 2.1} The function 
$f_\chi\circ\exp$ is Ad$_K$ invariant on the Lie algebra $\frak u_\bC$ of $U_\bC$.
\endproclaim

\demo{Proof}
For any $k, k_0\in K, \zeta\in\frak u_\bC$
$$
\chi(k\exp(-\text{Ad}(k_0)\zeta))=\chi(kk_0\exp(-\zeta)k^{-1}_0)=
\chi(k^{-1}_0kk_0\exp(-\zeta)).
$$
Thus 
$$
f_\chi(\exp(\text{Ad}(k_0)\zeta))=
\int\limits_{K}\chi(k^{-1}_0kk_0\exp(-\zeta))dk=
f_\chi(\exp(\zeta)).
$$
\qed
\enddemo

\proclaim{Proposition 2.2} Let $F\in\Cal O(\Bbb C)$ be an even function and
 $v\in \frak p_*$. Then  $F(\ad(v))$ (defined by its power series)
maps  $\bC\otimes\frak p_*$ into itself  and 
$\det(F(\left.\ad(v))\right|_{\bC\otimes\frak p_*})$ is an Ad$_K$ invariant
function.
\endproclaim

\demo{Proof} 
For every $k$ in $K$, Ad$(k)$ is in Aut$(\frak u)$. Thus for every 
$v\in \frak u$, $l=0,1,\dots$
$$
(\ad({\text{Ad}(k)v)})^l=\text{Ad}(k)\circ(\ad(v))^l\circ\text{Ad}(k)^{-1}.
$$
Hence
$$
F(\ad(\text{Ad}(k)v))=\text{Ad}(k)\circ F(\ad(v))\circ\text{Ad}(k)^{-1}.
$$
Since $\frak p_*$ is both Ad$(k)$ and $(\ad(v))^{2l}$ invariant ($l=0,1\dots$),
 the statement follows. 
\qed
\enddemo

From now on we shall assume that $M$ is a rank$-1$ symmetric space.

 Let $H_0\in\frak p_*$ with $\|H_0\|=1$. Then $\frak a_*=\bR H_0$ is maximal 
Abelian in $\frak p_*$ (resp. $\frak a:=i\frak a_*$ in $\frak p:=i\frak p_*$).
 Let $\Sigma$ be the set of restricted roots
corresponding to $(\frak g_0:=\frak k+\frak p, \frak a)$. Let 
$\frak a^+:=\{irH_0\mid r>0\}$ be the Weyl chamber and $\Sigma^+$ the set of 
positive restricted roots. Then $\Sigma^+=\{\beta,\beta/2\}$ with an apropriate
$\beta$ in the dual of $\frak p$ with $B:=\beta(iH_0)>0$. The corresponding
multiplicities are $m_\beta$ and $m_{\beta/2}$, where our convention is that
 the latter is zero when $\Sigma$ is reduced (i.e. when $M$ is a sphere with
the round metric).    

Let $\Bbb Z_+=\{0,1,2,\dots\}$.
According to Helgason's theorem ([He2, Theorem 4.1,(ii), p.535 
and Sect.3, p.542]), the set of linear functionals 
$\{\mu=n\beta: n\in\Bbb Z_+\}$ is precisely the set of 
    the highest weights of all irreducible spherical
representations of $U$ (w.r.t. $K$) restricted to $\frak a$.
 Now in light of 
what was said at the beginning of Sect.2 about the relationship of $V_\chi$ and
the irreducible spherical representations of $U$, we can conclude that
the isotypical subspaces $V_\chi$ of $\hat\pi$ are parametrized by the elements
 $n_\chi$ of $\Bbb Z_+$.
Let 
$$
a_\chi=\frac1{2}m_{\beta/2}+m_\beta+n_\chi,\quad b_\chi=-n_\chi,\quad c_\chi=
\frac{m_{\beta/2}+m_\beta+1}{2}=\frac{m}{2},\tag2-2
$$
and denote by $F_\chi$ the Gauss hypergeometric function,
 corresponding to these parameters
$$
F_\chi(x)=F(a_\chi,b_\chi,c_\chi,x).\tag2-3
$$
(see Sect.5 for more on hypergeometric functions).

Let $S^{m-1}_{\frak p_*}$ be the unit sphere in the euclidean space $\frak p_*$.

\proclaim{Theorem 2.3} Let $M^m$ be a  compact,  simply connected, rank-1 
symmetric space.  Then
$$
p_\chi(s)=
\frac{c_\chi\text{\rm Vol}(S^{m-1}_{\frak p_*})2^{\frac{m}{2}}}{B^m
(\text{Im s})^{\frac{m}{2}}}
\int\limits^\infty_0e^{-\frac{t^2}{B^2\text{Im }s}}
F_\chi(-\sh^2(t))t^{\frac{m-1}{2}}(\sh(t))^{\frac{m-1}{2}}
(\ch(t))^{\frac{m_\beta}{2}}dt.\tag2-4
$$
\endproclaim

\demo{Proof}
Suppose $f\in L^1(\frak p_*)$  depends only on $\|\zeta\|$.
Using polar coordinates we obtain
$$
\int\limits_{\frak p_*}f(\zeta)d\zeta=\text{\rm Vol}(S^{m-1}_{\frak p_*})
\int\limits^\infty_0f(rH_0)r^{m-1}dr.\tag2-5
$$
According to 
\cite{He2, formula (25), p.543}, the spherical function $f_\chi$ 
can be expressed as 
$$
f_\chi(\exp(2irH_0))=F_\chi(-\sh^2(\beta(irH_0)))=
F_\chi(-\sh^2(rB)).\tag2-6
$$

Prop.2.2 applied  to  $F(z)=\frac{\sin2z}{z}$ shows that $\eta$ is Ad$_K$ 
invariant. Since the rank is $1$,  Ad$_K$ acts transitively on each sphere with
center the origin in $\frak p_*$. In light of Prop.2.1,  
(2-5)  applies to the integrand in (2-1) and we get
$$
p_\chi(s)=
\frac{c_\chi\text{\rm Vol}(S^{m-1}_{\frak p_*})}{
(\text{Im }s)^{\frac{m}{2}}}
\int\limits^\infty_0e^{-\frac{r^2}{\text{Im }s}}
F_\chi(-\sh^2(rB))r^{m-1}\sqrt{\eta}(rH_0)dr.\tag2-7
$$
For $H\in\frak a_*$,
$\ad H^2:\frak p_*\rightarrow\frak p_*$ has eigenvalues 
$0$ with multiplicity $1$, $\beta(H)^2$ with multiplicity $m_\beta$ and
$(\beta(H)/2)^2$ with multiplicity $m_{\beta/2}$ (\cite{He1, Lemma 2.9, p288}). 
Thus
$$
\eta(H)=2\left(\frac{\sin(2\beta(H))}{\beta(H)}\right)^{m_\beta}
\left(\frac{\sin(\beta(H))}{\beta(H)/2}\right)^{m_{\beta/2}}.\tag2-8
$$
Now for $H=rH_0$  we have $\beta(rH_0)=-irB$. Hence (2-8) yields
$$
\eta(rH_0)=\frac{2^m}{(rB)^{m-1}}
(\sh(rB))^{m-1}
(\ch(rB))^{m_{\beta}}.
$$
Substituting this into (2-7) and changing the variable $r$ in the integral to
$t=rB$ we finally get formula (2-4).
\qed
\enddemo
From   (2-4) we see that  $p_\chi(s)$  depends only on $\tau=B^2$Im$s$. In light
of our earlier characterization of the (projective) flatness of $H^{corr}\ra S$
in term of $\bar\partial\partial\log p_\chi$
 we obtain:
\proclaim{Corollary 2.4} Let $M$ be a compact, simply connected rank-1 
symmetric space. 
For $\tau>0$ let 
$$
q_\chi(\tau):=\int\limits^\infty_0e^{-\frac{t^2}{\tau  }}
F_\chi(-\sh^2(t))t^{\frac{m-1}{2}}(\sh(t))^{\frac{m-1}{2}}
(\ch(t))^{\frac{m_\beta}{2}}dt.\tag2-9
$$
Then the field  of quantum Hilbert spaces $H^{corr} \rightarrow S$
is
$$
\aligned
\text{ flat iff}\quad (\log q_\chi(\tau))''\equiv0\quad \text{for every } 
\chi,\\
\text{ projectively flat iff}\quad(\log q_\chi(\tau))''\quad
\text{does not depend on }  \chi.\endaligned\tag2-10
$$
\endproclaim
The integral in (2-9) can be explicitely calculated only in very special cases.
To be able to decide whether (2-10) holds, 
we shall use asymptotic methods to investigate
the behavior of $q_\chi(\tau)$ as $\tau\rightarrow 0$ and 
$\tau\rightarrow\infty$.
As we will see in Sect.5, the function $F_\chi(x)$ is a polynomial of degree 
$n_\chi$. This motivates our investigations in the next sections.

\subhead 3. $Q_P$ functions and central polynomial sequences
\endsubhead

Let  $P(x)=c_nx^n+\dots+c_1x+c_0$ be a polynomial ($c_j\in \Bbb C$), 
$\tau>0$ and
 $  \mu,\kappa, \nu\in\Bbb C,$ with  Re$(\mu+\kappa)>-1$.
Define the corresponding $Q_P$ function by the formula
$$
Q_P(\tau):=\int\limits^\infty_0e^{-\frac{t^2}{\tau}}P(-\sh^2t)
t^\mu(\sh t)^\kappa(\ch t)^\nu dt.\tag3-1
$$
The integral converges absolutely  and $Q_P$ depends holomorphically 
  on the
parameters $ \mu,\kappa,\nu$.
The function
$$
f_P(t):=P(-\sh^2t)\left(\frac{\sh t}{t}\right)^\kappa (\ch(t))^\nu\tag3-2
$$
is  even and extends  
 holomorphically to  a neighborhood of the real line. 
Let  $r:=\kappa+\mu+1.$
Then
$$
Q_P(\tau)=\int\limits^\infty_0e^{-\frac{t^2}{\tau}}t^{r-1}f_P(t)dt.\tag3-3
$$
Applying Watson's  lemma [W] to this integral we get
$$
Q_P(\tau)=\frac{\tau^{\frac{r}{2}}}{2}
\left(\Gamma\left(\frac{r}{2}\right)f_P(0)+
\Gamma\left(\frac{r}{2}+1\right)\frac{f''_P(0)}{2}\tau+
o(\tau)\right),
\quad \tau\rightarrow 0,
$$
where $\Gamma$ denotes the usual gamma function.
Now $f_P(0)=c_0$ and a  straightforward calculation shows 
$$
\frac{f''_P(0)}{2}=-c_1+\frac{\kappa}{6}+\frac{\nu}{2},
$$
and we get:
\proclaim{Proposition 3.1}
$$
Q_P(\tau)=\frac{\tau^{\frac{r}{2}}}{2}\left(\Gamma\left(\frac{r}{2}\right)c_0+
\Gamma\left(\frac{r}{2}+1\right)\left(-c_1+
\frac{\kappa}{6}+\frac{\nu}{2}\right)\tau+
o(\tau)\right),
\quad \tau\rightarrow 0, 
$$
\endproclaim
The next definition is motivated by Corollary 2.4.
\definition{Definition 3.2}
 Let $\{P_n(x)=c_{n,n}x^n+\dots+c_{n,1}x+1\}^\infty_{n=0}$, $c_{n,n}\not=0$ 
($c_{n,j}\in\Bbb C$)
 be a sequence of polynomials. 
The sequence is
 called 
{\it central} (w.r.t. the parameters $\mu,\kappa,\nu$) if the function
 $(\log{Q_{P_n}})''$ does not depend on $n$. 
\enddefinition

\proclaim{Proposition 3.3} Suppose $\{P_n\}^\infty_{n=0}$ is a 
{\it central} sequence of polynomials. Then 

$$
Q_{P_n}(\tau)=e^{-\frac{r}{2} c_{n,1}\tau}Q_1(\tau),\quad  n\ge1,\quad
\text{where }r=\mu+\kappa+1. \tag3-4
$$
\endproclaim

\demo{Proof} From our assumption 
$$
\left(\log Q_{P_n}-\log Q_{P_0}\right)''\equiv0,
$$
for every $n$. Hence there exist
 constants $\alpha_n$ and  $\beta_n$  such that
$$
Q_{P_n}(\tau)=\beta_ne^{\alpha_n\tau}Q_{P_0}(\tau)\tag3-5
$$
Substituting this into the asymptotic formula in Prop.3.1 we get
$$\aligned
\beta_n(1+\alpha_n\tau+o(\tau))&\frac{\tau^{\frac{r}{2}}}{2}
\left(\Gamma\left(\frac{r}{2}\right)+
\Gamma\left(\frac{r}{2}+1\right)\left(\frac{\kappa}{6}+\frac{\nu}{2}\right)\tau+
o(\tau)\right)=\\
&=\frac{\tau^{\frac{r}{2}}}{2}\left(\Gamma\left(\frac{r}{2}\right)+
\Gamma\left(\frac{r}{2}+1\right)\left(-c_{n,1}+
\frac{\kappa}{6}+\frac{\nu}{2}\right)\tau+
o(\tau)\right).
\endaligned\tag3-6
$$
Now dividing  by $\frac{\tau^{\frac{r}{2}}}{2}$ 
and comparing  the constant term 
and the 
coefficient of $\tau$ on both sides of 
(3-6) we get
$$
\beta_n=1,\tag3-7
$$
and 
$$
\beta_n\alpha_n\Gamma\left(\frac{r}{2}\right)+
\Gamma\left(\frac{r}{2}+1\right)\beta_n\left(\frac{\kappa}{6}+
\frac{\nu}{2}\right)
=
\Gamma\left(\frac{r}{2}+1\right)\left(-c_{n,1}+\frac{\kappa}{6}+
\frac{\nu}{2}\right)
.\tag3-8
$$
 Hence $\alpha_n=-rc_{n,1}/2.$
\qed
\enddemo

\subhead 4. Asymptotics at infinity
\endsubhead
\proclaim{Proposition 4.1} Let
$a\ge0, \lambda>0,\mu>-1, \tau>0.$ Then
$$
I(\tau):=\int\limits^\infty_ae^{-t^2/\tau}t^\mu e^{\lambda t}dt=
\frac{\lambda^\mu\sqrt{\pi}}{2^\mu}\tau^{\mu+1/2}e^{\lambda^2\tau/4}(1+o(1)),
\qquad   
\tau\longrightarrow\infty.
$$
\endproclaim

\demo{Proof}
We can rewrite $I(\tau)$ as follows:
$$
I(\tau)=e^{\lambda^2\tau/4}\int\limits^\infty_ae^{-(\frac{t}{\sqrt{\tau}}-
\frac{\lambda\sqrt{\tau}}{2})^2}t^\mu dt.
$$
After the substitution $x=\frac{t}{\sqrt{\tau}}-
\frac{\lambda\sqrt{\tau}}{2}$, 
we get

$$
I(\tau)=\tau^{\mu+1/2}e^{\lambda^2\tau/4}\int\limits^{+\infty}_
{\frac{a}{\sqrt{\tau}}-
\frac{\lambda\sqrt{\tau}}{2}} 
e^{-x^2}\left(\frac{x}{\sqrt{\tau}}+\frac{\lambda}{2}\right)^\mu dx=:
\tau^{\mu+1/2}e^{\lambda^2\tau/4}I_2(\tau).
$$
 Lebesgue's dominated convergence theorem applied to
  $I_2(\tau)$ yields:

$$
I_2(\tau)\longrightarrow\left(\frac{\lambda}{2}
\right)^\mu\int\limits^{+\infty}_{-\infty}
e^{-x^2}dx=\left(\frac{\lambda}{2}
\right)^\mu\sqrt\pi,\qquad \tau\rightarrow \infty,
$$
finishing the proof.
\qed
\enddemo

\proclaim{Proposition 4.2} Let $\nu >0, \kappa>0, \mu+\kappa>-1$ $\tau>0$.  Then
$$
\int\limits^\infty_0e^{-\frac{t^2}{\tau}}t^\mu(\sh t)^\kappa(\ch t)^\nu dt=
\frac{\sqrt\pi(\nu+\kappa)^\mu}{2^{\mu+\nu+\kappa}}\tau^{\mu+\frac1{2}}
e^{\frac{(\kappa+\nu)^2}{4}\tau}
(1+o(1)),\quad
 \tau\rightarrow\infty.
$$
\endproclaim

\demo{Proof} Let $a>0$ be an arbitrary fixed constant. Then
$$
\int\limits^a_0e^{-\frac{t^2}{\tau}}t^\mu(\sh t)^\kappa(\ch t)^\nu dt=O(1)=
\tau^{\mu+\frac1{2}}
e^{\frac{(\kappa+\nu)^2}{4}\tau}o(1),\quad \tau\rightarrow\infty.\tag4-1
$$
Let  $h(x)=(1-x)^\kappa(1+x)^\nu$. Then $h\in C^\infty(-1,1)$, $h(0)=1$. Hence 
$h(x)=1+xg(x)$, where $g\in C^\infty(-1,1)$. Therefore
$$
(\sh t)^\kappa(\ch t)^\nu=
\frac{e^{(\kappa+\nu)t}}{2^{\kappa+\nu}}(1-e^{-2t})^\kappa(1+e^{-2t})^\nu=
\frac{e^{(\kappa+\nu)t}}{2^{\kappa+\nu}}(1+e^{-2t}g(e^{-2t})).
$$
Thus
$$\aligned
\int\limits^\infty_ae^{-\frac{t^2}{\tau}}t^\mu(\sh t)^\kappa(\ch t)^\nu dt&=\\
=\frac1{2^{\kappa+\nu}}\int\limits^\infty_ae^{-\frac{t^2}{\tau}}
t^\mu e^{(\kappa+\nu)t}&dt+
\frac1{2^{\kappa+\nu}}\int\limits^\infty_ae^{-\frac{t^2}{\tau}}t^\mu 
e^{(\kappa+\nu-2)t}
g(e^{-2t})dt=I_1+I_2.
\endaligned\tag4-2
$$
Now by Proposition 4.1
$$
I_1=\frac{\sqrt\pi(\nu+\kappa)^\mu}{2^{\mu+\nu+\kappa}}\tau^{\mu+\frac1{2}}
e^{\frac{(\kappa+\nu)^2}{4}\tau}
(1+o(1)).\tag4-3
$$
The function
 $g(e^{-2t})$ is bounded on $[a,\infty)$, 
thus for an appropriate constant $A$ we
have
$$
|I_2|\le A\int\limits^\infty_0
e^{-\frac{t^2}{\tau}}t^\mu e^{(\kappa+\nu-2)t}dt=:AI_3.
$$ 
If $\kappa+\nu-2>0$, by Proposition 4.1
$$
I_3=\tau^{\mu+\frac1{2}}
e^{\frac{(\kappa+\nu-2)^2}{4}\tau}O(1)=\tau^{\mu+\frac1{2}}
e^{\frac{(\kappa+\nu)^2}{4}\tau}o(1).
$$
 If $\kappa+\nu-2\le0$, after the substitution $t=\sqrt{\tau x}$ we obtain
$$
I_3\le \int\limits^\infty_0
e^{-\frac{t^2}{\tau}}t^\mu dt=\frac{\tau^{\frac{\mu+1}{2}}}{2}\int\limits^\infty_0
e^{-x}x^{\frac{\mu-1}{2}}dx=\frac{\tau^{\frac{\mu+1}{2}}}{2}
\Gamma\left(\frac{\mu+1}{2}\right).
$$

Thus in both cases 

$$
I_2=\tau^{\mu+\frac1{2}}
e^{\frac{(\kappa+\nu)^2}{4}\tau}o(1).\tag4-4
$$
Then (4-1), (4-2), (4-3) and (4-4) together prove our claim.
\qed
\enddemo

\proclaim{Proposition 4.3} Let $\nu>0,\kappa>0,$ $\mu+\kappa>-1$ and 
$P(t)=c_nt^n+\dots+c_0$ be a polynomial. Let $Q_P$ be 
the corresponding $Q$ function (see (3-1)).
Then
$$Q_P(\tau)=(-1)^nc_n\frac{\sqrt\pi(\nu+\kappa+2n)^\mu}{2^{\mu+\nu+\kappa+2n}}
\tau^{\mu+\frac1{2}}
e^{\frac{(\kappa+\nu+2n)^2}{4}\tau}
(1+o(1)),\quad
 \tau\rightarrow\infty.
$$
\endproclaim
\demo{Proof}
It is enough to show the statement for the special case $P(t)=t^k$. Then
$$
Q_P(\tau)=
\int\limits^\infty_0e^{-\frac{t^2}{\tau}}(-\sh^2t)^kt^\mu(\sh t)^\kappa(\ch t)^\nu dt=
(-1)^k\int\limits^\infty_0e^{-\frac{t^2}{\tau}}t^\mu(\sh t)^{\kappa+2k}(\ch t)^\nu dt
$$ 
and Proposition 4.2 proves our claim.
\qed
\enddemo

\proclaim{Theorem 4.4} Let $\nu>0,\kappa>0, \mu+\kappa>-1$. Suppose
 $\{P_n\}^\infty_{n=0}$ is a 
central sequence of polynomials.
Then  for all $n$
$$
Q_{P_n}(\tau)=e^{n(\nu+\kappa+n)\tau}Q_1(\tau),\tag4-5
$$
$$
c_{n,1}=-\frac{2n(\nu+\kappa+n)}{\mu+\kappa+1},\quad
c_{n,n}=(-1)^n4^n\left(\frac{\nu+\kappa}{\nu+\kappa+2n}\right)^\mu.\tag4-6
$$
\endproclaim
\demo{Proof}
Centrality implies (see (3-4)) that $Q_{P_n}(\tau)=A_ne^{B_n\tau}Q_1(\tau)$ 
for appropriate
constants $A_n, B_n$. Substituting into this the asymptotics of 
Proposition 4.3 and comparing the leading terms on both sides we get
$$
c_{n,n}(-1)^n\frac{(\nu+\kappa+2n)^\mu}{4^n}e^{\frac{(\nu+\kappa+2n)^2}{4}\tau}=
A_ne^{B_n\tau}(\nu+\kappa)^\mu e^{\frac{(\nu+\kappa)^2}{4}\tau}.
$$
Therefore
$$
B_n=\frac{(\nu+\kappa+2n)^2-(\nu+\kappa)^2}{4}=n(\nu+\kappa+n)
$$
and
$$
A_n=c_{n,n}(-1)^n\left(\frac{\nu+\kappa+2n}{\nu+\kappa}\right)^\mu\frac1{4^n}.
$$
Now comparing these expressions of $A_n, B_n$ with (3-4) we get (4-5) and 
(4-6). 
\qed
\enddemo
Taking $n=1$ in (4-6)  we get two different expressions for the same
coefficient and thus we obtain the following.
\proclaim{Corollary 4.5} Let $\nu,\kappa>0, \mu+\kappa>-1$. 
Suppose there exists a 
 central sequence of polynomials corresponding to these parameters.
Then
$$
\frac{\nu+\kappa+1}{\mu+\kappa+1}=
2\left(\frac{\nu+\kappa}{\nu+\kappa+2}\right)^\mu.
$$
\endproclaim

\subhead 5. Hypergeometric polynomials and the proof of Theorem 1.1
\endsubhead

Recall that Gauss' hypergeometric functions 
are given by


$$
F(a,b,c,z):=1+\frac{ab}{c}z+\dots
+\frac{a(a+1)\dots(a+k-1)b(b+1)\dots(b+k-1)}{k!c(c+1)\dots(c+k-1)}
z^k+\dots\tag5-1
$$
where $a,b,c\in\Bbb C$, $c\not\in \Bbb Z_{-}=\{0,-1,-2,\ldots\}$. The series
 converges at least in the unit disk. 
If $n\in\Bbb Z_+=\{0,1,2,\ldots\}$, $b=-n$, $A\in\Bbb C\setminus\Bbb Z_-$,  and 
$a=A+n$,  then  $F$  is a
polynomial (in $z$)  of degree $n$. 
Now assume $A, c\in \Bbb R\setminus \Bbb Z_{-}$, $n\in\Bbb Z_+$ and  
consider this 
 sequence of polynomials
$$
P_n(x):=F(A+n,-n,c,x)=\sum\limits^n_{j=0}c_{n,j}x^j.\tag5-1a
$$

\proclaim{Proposition 5.1} Suppose the polynomial sequence $P_n(x)$ is central
 w.r.t.  some choice of the  parameters $\nu, \kappa, \mu$ with 
$\nu>0, \kappa>0,\mu+\kappa>-1$. Then
$A>0, c>0$,
$$
\frac{\Gamma(A+2n)}{\Gamma(A+n)}\frac{\Gamma(c)}{\Gamma(c+n)}=
4^n\left(\frac{A}{A+2n}\right)^\mu,\tag5-2
$$
and
$$
\kappa=2c-\mu-1,\quad \nu=A-\kappa.
\tag5-3
$$  
\endproclaim
\demo{Proof} Assume $\{P_n\}^\infty_0$ is central.
Then (5-1) and  (4-6) yield
$$
c_{n,1}=\frac{(A+n)(-n)}{c}=-\frac{2n(\nu+\kappa+n)}{\mu+\kappa+1},
$$
for every $n$, which implies
$$
c=(\mu+\kappa+1)/2,\quad A=\nu+\kappa,\tag5-4
$$
proving (5-3) and $A, c>0$.
 From (5-1) we also get  
$$
c_{n,n}
=(-1)^n\frac{\Gamma(A+2n)}{\Gamma(A+n)}\frac{\Gamma(c)}{\Gamma(c+n)},
$$
which together with  formula (4-9) and (5-4) proves (5-2).
\qed
\enddemo

\subhead{Proof of Theorem 1.1}
\endsubhead
 Suppose the field $H^{corr}$  of quantum Hilbert 
spaces corresponding to $M$ is projectively flat. As in Section 2, with $M$ 
we associate its system of restricted roots $\Sigma$. 
We denote the longer 
positive restricted root $\beta$, and the multiplicities of 
$\beta$, $\beta/2$ by $m_\beta$, $m_{\beta/2}$ 
(with the understanding that this latter is $0$ if $\Sigma$ 
is reduced,  so that 
$\beta/2$ is not a root, i.e. when
 $M$ is a sphere). Then $m=m_\beta+m_{\beta/2}+1$.
Set
$$
 A=m_\beta+\frac{m_{\beta/2}}{2},\quad c=\frac m2,
$$ 
and $P_n(x)=F(A+n,-n,c,x)$, $n=1,2,\ldots$, $P_0(x)=1$ 
the corresponding sequence of 
hypergeometric polynomials. Let also $\mu=\kappa=(m-1)/2$, $\nu=m_{\beta}/2$.
According to Corollary 2.4 and  Definition 3.2, projective flatness implies 
that  $P_n$ is central with respect to $\kappa,\mu,\nu$.
 
We now apply 
Proposition 5.1.
 Choose $n=2A$, so that $A/(A+2n)=1/5$. The  left hand side 
of (5-2) is rational, hence $\mu=(m-1)/2$ on the right must be an integer and 
so $m$ must be odd.  Then it follows
from the classification of compact rank-1 symmetric spaces
 (see [He2, Ch.I, Sect.4.2]),
 that $M$ must be an odd dimensional sphere and so  $\Sigma$ is reduced. 
Thus $m_{\beta/2}=0$, and $A=m_\beta=m-1$.
Substitute $n=1$ into (5-2):
$$ 
2=\left(\frac{m+1}{m-1}\right)^{\frac{m-1}{2}}=\left(1+
\frac 2{m-1}\right)^{\frac{m-1}{2}}.
$$
Clearly $m=3$ solves this equation, and there is no other solution, because 
$(1+1/x)^x$ is strictly increasing for $x>0$.
Thus $M$ must be the 3 dimensional sphere: for no other compact simply 
connected symmetric space of rank $1$ can the Hilbert field $H^{corr}$ be 
projectively flat.

On the other hand, when $M$ is $S^3$, or more generally a compact Lie group 
with biinvariant metric, the associated Hilbert field is outright flat, 
see [L-Sz, Theorem 11.3.1].\qed

\enddocument
\bye